# Generating nanoscale and atomically-sharp p-n junctions in graphene via monolayer-vacancy-island engineering of Cu surface

Ke-Ke Bai[1,§], Jiao-Jiao Zhou[2,§], Yi-Cong Wei[1], Jia-Bin Qiao[1], Yi-Wen Liu[1], Hai-Wen Liu[1], Hua Jiang[2,*], Lin He[1,*]

[1]Center for Advanced Quantum Studies, Department of Physics, Beijing Normal University, Beijing, 100875, People's Republic of China

[2]College of Physics, Optoelectronics and Energy, Soochow University, Suzhou, 215006, People's Republic of China

[§]These authors contributed equally to this work.

[*]Correspondence and requests for materials should be addressed to H. J. (e-mail: jianghuaphy@suda.edu.cn), and L. H. (e-mail: helin@bnu.edu.cn).

**Creation of high quality *p-n* junctions in graphene monolayer is vital in studying many exotic phenomena of massless Dirac fermions. However, even with the fast progress of graphene technology for more than ten years, it remains conspicuously difficult to generate nanoscale and atomically-sharp *p-n* junctions in graphene. Here, we employ monolayer-vacancy-island engineering of Cu surface to realize nanoscale *p-n* junctions with atomically-sharp boundaries in graphene monolayer. The variation of graphene-Cu separations around the edges of the Cu monolayer-vacancy-island affects the positions of the Dirac point in graphene, which consequently lead to atomically-sharp *p-n* junctions with the height as high as 660 meV in graphene. The generated sharp *p-n* junctions isolate the graphene above the Cu monolayer-vacancy-island as nanoscale graphene quantum dots (GQDs) in a continuous graphene sheet. Massless Dirac fermions are confined by the *p-n* junctions for a finite time to form quasi-bound states in the GQDs. By using scanning tunneling microscopy, we observe resonances of quasi-bound states in the GQDs with various sizes and directly visualize effects of geometries of the**

**GQDs on the quantum interference patterns of the quasi-bound states, which allow us to test the quantum electron optics based on graphene in atomic scale.**

Electronic junctions (*p-n* junctions) in graphene have attracted much attention over the years not only because of their importance in fundamental science, for example in studying the Klein tunneling of massless Dirac fermions[1-6], but also because they will be essential components in quantum electron optics based on graphene[7-15]. Recent experiments demonstrated that it is possible to manipulate electron refraction and transmission in graphene like photons by using *p-n* junction[11,13-15], in which the width and the sharpness of the junctions play crucial roles in precisely manipulating electrons. Therefore, generating high-quality *p-n* junctions in the nanometer regime and, simultaneously, with atomically-sharp boundaries in graphene is one of the central goals of graphene-based nanoelectronics. However, such a goal has been demonstrated to be quite difficult to achieve[14-17] and seems to be not within the grasp of today's technology. Until very recently, substrate engineering exhibits the ability to create sharp potential wells in graphene[18,19], which makes it possible to manipulate the massless Dirac fermions in the same way as lights. In this work, we demonstrate an approach based on monolayer-vacancy-island engineering of Cu substrate to generate nanoscale and atomically-sharp junctions in a continuous graphene sheet. Taking advantage of the monolayer-vacancy-island on Cu surface, the separations between the graphene sheet and the Cu substrate vary sharply around the edges of the vacancy-island, which naturally affects positions of Dirac points in graphene and generates atomically-sharp *p-n* junctions with the height as high as about 660 meV. This provides us unprecedented opportunity to directly image the wavefunctions around the *p-n* junctions and atomically verify the quantum electron optics based on graphene.

Previous studies demonstrated that reconstructions of metals, such as Cu, Ag and Au, could induce both islands and vacancy-islands on the surface[20-24]. In this work, monolayer-vacancy-islands on Cu surface were generated during the synthesis of the graphene. In our experiment, graphene monolayer was grown on Cu foils by chemical vapor deposition (CVD) method (see Supplemental Material for details)[25]. The Cu foil was annealed in 1030 ℃ at atmospheric pressure for several hours to form large scale single-crystal Cu surface (mainly exposed Cu (111) surface, see in supplemental Fig. S1), and then graphene monolayer was grown on it by the low pressure CVD method. During the high-temperature annealing process, the surface Cu atomic vacancies and adatoms emerge through migration and massive diffusion of the surface Cu atoms[21-24], as schematically shown in Fig. 1a. The as-grown graphene sheet (with the lattice constant 0.246 nm) will introduce a local stress field in the Cu surface (with the nearest-neighbor atomic spacing ∼ 0.256 nm), which is expected to reduce the activation barrier for the diffusion of the Cu atomic vacancies[20]. Therefore, we observed many Cu monolayer-vacancy-islands, with the sizes ranging from several nanometers to dozens of nanometers, on the synthesized samples, as shown in Fig. 1b as an example. The graphene-Cu separation above the monolayer-vacancy-islands, $d_2$, is expected to be larger than that on other parts of the Cu surface, $d_1$, as schematically shown in the lower panel of Fig. 1a. We will demonstrate subsequently that the atomic-layer difference of graphene-Cu separations between inside and outside of the vacancy-islands can introduce sharp electronic junctions in the continuous graphene monolayer.

Figure 1b shows a representative large-area scanning tunneling microscopy (STM) image of the graphene monolayer on the Cu substrate. We can observe several protrusions in the STM image. Fig. 1c shows an enlarged STM image of a protrusion shown in Fig. 1b. Obviously, the graphene sheet in the entire area of Fig. 1c is continuous and free of defects. The high density of atomic depressions (~20-30 pm in depth), peppering the surface surrounding the protrusion, are atomic vacancies of the Cu surface. The zoom-in atomic-resolution of graphene on the Cu surface with atomic-vacancy-defects is shown in Fig. S2 in the supplemental material. The height profiles across the protrusion recorded at different biases vary a lot, as shown in Fig. 1d, indicating that the protrusion in the STM image is mainly contributed by the electronic effects (see more height profiles of the GQD at different biases in supplemental Fig. S3). Here we attribute the protrusions to the effects of Cu vacancy-islands on the electronic states of the suspended graphene sheet. During the synthesis process of graphene, the Cu atomic vacancies can merge into nanoscale vacancy-islands on Cu surface. The variations of graphene-Cu separations inside and outside of the vacancy-islands affect the overlap of graphene and Cu wave functions, which consequently dope the graphene differently[26].

To verify the above assumption, we carried out both field-emission resonances (FER) and scanning tunneling spectroscopy (STS) measurements of graphene on and off the vacancy-island region. According to the energy shifts in the FER peaks[18], we directly obtain the differences in the local work function of graphene on and off the vacancy-island. The changes of the local work function in our experiment are measured to be about 680 meV (see supplemental Fig. S4 for the details). Such a result is further confirmed in our STS measurements. Fig. 1e displays two representative STS curves recorded inside and outside the vacancy-island. The STS spectrum gives direct access to the local density of states (LDOS) of the surface underneath the STM tip. A local dip in the tunneling spectra, as pointed out in Fig. 1e, reflects the vanished LDOS of graphene at the Dirac point $E_D$[27-30]. The Dirac point recorded within the vacancy-island is located at about 350 meV, indicating *p*-type doping; while the Dirac point recorded outside the vacancy-island is located at about -310 meV, indicating *n*-type doping. Obviously, the obtained energy difference of the Dirac points, ~ 660 meV, consists well with that obtained by the FER measurements. Similar observations are obtained around tens of the vacancy-islands in our experiment. These results indicate that variations of graphene-Cu separations inside and outside of the vacancy-islands generate $p$-$n$ junctions with potential barrier as high as about 660 meV in graphene. Fig. 1f summarizes the energies of the Dirac points $E_D$, acquired from the STS curves, as a function of positions along the magenta arrow in Fig. 1c. Obviously, the transition from the $n$-type to the $p$-type doping occurs sharply within 1 nm. With considering $E_D \sim$ -310 meV recorded outside the vacancy-island and assuming that the Dirac point of graphene is only determined by the graphene-Cu separation[26], we obtained the graphene-Cu separation outside the vacancy-island as about $d_1 = 0.32$ nm ($d_1$ is defined in Fig. 1a. See supplemental (A.3) and Fig. S5 for the detail of calculation), which agrees with the equivalent distance between graphene and Cu (111) surface. Similarly, we obtained the average graphene-Cu separation within the vacancy-islands as about $d_2 = 0.50$ nm. The difference between $d_1$ and $d_2$ is about 0.18 nm (Fig. 1f), almost the same as the interlayer spacing between the Cu (111) planes, indicating that the vacancy-islands on the Cu surface are monolayer-vacancy-islands. Additionally, the more apparent and easily obtained atomic-resolution STM images of the graphene above the vacancy-islands indicates the weaker coupling between graphene and Cu substrate in this region than that off the vacancy-islands, which further verifies the larger graphene-Cu separation on the vacancy-islands than that off the vacancy-islands.

The atomically-sharp $p$-$n$ junctions along the edges of the monolayer-vacancy-islands isolate the graphene above the islands as nanoscale and atomically-sharp graphene quantum dots (GQDs). Because of the unusual anisotropic transmission of the massless Dirac fermions at the $p$-$n$ junction, i.e., the Klein tunneling, quasi-particles in graphene incident at large oblique angles will be reflected from the junction with high probability[1-6]. These reflected quasi-particles are trapped in the GQDs with finite trapping time and form quasi-bound states in the GQDs[31-35]. The emergence of the resonant peaks below the Dirac point in the tunneling spectra, as shown in Fig. 1e, is a clearly evidence of the formation of quasi-bound states in the GQDs[14,17-19].

Based on the monolayer-vacancy-islands engineering of Cu surface, it is facile to obtain the GQDs with different sizes and geometries embedded in a continuous graphene sheet. Fig. 2 shows several typical GQDs obtained in our experiment. These GQDs with various geometries provide us unprecedented opportunities to directly visualize the effects of the geometry on the interference patterns of the quasi-bound states, which can be treated as atomic-scale verification of the quantum electron optics in graphene.

Fig. 3a shows four representative STS spectra recorded inside the GQDs with different sizes and geometries. Almost equally spaced resonances, ascribed to the quasi-bound states[14,17-19], are observed below the Dirac points in all the spectra. Since the GQDs in our experiment have quite different geometries, we define the effective radius of the GQDs via $R = \sqrt{A/\pi}$ for simplicity, where $A$ is the area of each GQD measured in STM images. The STS spectra shown in Fig. 3a indicate that the average level spacing for the resonant peaks decreases with the effective radius $R$. For massless Dirac fermions confined in a circular GQD with the radius $R$, the average level spacing of the resonant states can be expressed as $\Delta E \approx \alpha \hbar v_F / R$, where $\alpha$ is a dimensionless constant of order unity, $\hbar$ is the reduced Planck's constant, and $v_F = 1.0 \times 10^6$ m/s is the Fermi velocity[18]. In Fig. 3b, we summarize the average level spacing of the resonant peaks ($\Delta E$) as a function of the inverse effective radius ($1/R$) of the GQDs measured in our experiment. The data can also be described well by $\Delta E \approx \alpha \hbar v_F / R$ with $\alpha = 0.96 \pm 0.06$, which indicates that the relation between the average level spacing of the resonant peaks and the size of the GQDs is insensitive to the geometry of the GQDs.

To further explore the electronic properties of the quasi-bound states in the GQDs, we carried out theoretical studies based on the lattice Green's function (see supplemental (A.4) for the calculated method). Considering the potential height and the sharpness of the *p-n* junction determined in experiment, the theoretical LDOS curves of circular GQDs (see supplemental Fig. S6d) capture well the main features of the experimental STS spectra. Furthermore, we also calculate a series of LDOS curves for circular GQDs (and several rectangle GQDs) with different size and potential height (see supplemental Fig. S7). Obviously, the calculated relation between $\Delta E$ and $1/R$ (obtained from the calculated LDOS curves in supplemental Fig. S7c), as plotted in Fig. 3b, can also be described well by $\Delta E \approx \alpha \hbar v_F / R$ and agrees quite well with our experimental result. The consistency between our experiment and simulation not only confirms the formation of quasi-bound states in the GQDs embedded in the continuous graphene sheet, but also demonstrates that the spectra of the quasi-bound states are weakly depending on the geometry of the GQDs.

Further inspection of the quasi-bound states trapping in the GQDs can be measured by STS maps, which reflect the spatial distribution of the LDOS (and consequently the spatial distribution of the confined massless Dirac fermions) at the recorded energies[36-38]. Fig. 4 shows several STS maps of two typical GQDs with different geometries obtained in our experiment (see supplemental Fig. S8 for more STS maps of GQDs with other different geometries). Obviously, the two GQDs exhibit quite different features in the STS maps. For the quasi-bound states formed in the quasi-circular GQD, the radial part of the wavefunction can be described by Bessel functions of the first kind[18]. Therefore, the wavefunction of the lowest resonant peak exhibits a maximum in the center of the GQD, and higher-energy resonances display shell structures with the maxima progressively approaching the edge of the GQD, as shown in Fig. 4a. However, for the quasi-bound states trapped in the quasi-rectangular GQD, they exhibit much more complex patterns with alternately dark and bright dots, as shown in Fig. 4b. The complex patterns originate from the constructive or destructive interference of the quasi-particles confined by the sharp junction boundaries. To further study effects of geometry of the GQDs on the quasi-bound states, we also carried out the Green's function simulation in diverse geometries of the GQDs, as shown in Fig. 4 for examples (see supplemental material (A.4) for details of calculation). Even though the roughness and irregularities of the edges of the GQDs are not taken into account in the calculation, the simulated LDOS maps reproduce quite well the main features of the STS maps obtained in our experiment. Our studies demonstrate that the geometry of the GQD has significant effect on the interference patterns of the quasi-bound states.

Although all our experimental results can be understood accurately by directly solving the Dirac equations or carrying out Green's function simulation in diverse geometries of the GQDs, we will show below that these results can also be explained by analogizing the GQDs with optical systems, which provides a more intuitive and simple physical picture. According to the Klein tunneling in graphene[1-6], massless Dirac fermions incident at large oblique angles can be reflected with high probability on the boundary of *p-n* junction. During such processes, the electrons behave like light. As shown in Fig. S9 in the supplemental material, after multiple reflections, electrons may return to the original place. If the dynamic phase difference satisfies $kL = 2\pi(n + \gamma)$, the constructive interference between the incident and reflected electrons form standing waves, leading to quasi-bound states. Here, k denotes the wave vector, L denotes length of closed interference path (see supplemental Fig. S9), $\gamma$ is the associate Berry phase in a closed path and n is an integer. Considering the linear dispersion of graphene, the peaks emerge at energy $E = 2\pi\hbar v_F(n + \gamma)/L$. This explains the almost equal-space resonant peaks in both Fig. 1e and Fig. 3a.

Next, we analyze the interference path in diverse geometries. For a circular GQD, there is only one incident angle $\theta$ for a quasi-particle incident to the junction boundary (see supplemental Fig. S9b). At larger $\theta$, the bound states become more confined due to the higher reflection probability[1] (see supplemental Fig. S9a-9c). This is the reason why the STS maps exhibit shell structures in quasi-circular GQDs (see Fig. 4a). Moreover, at large $\theta$, the length of path L approaches the circumference of GQD: $2\pi R$. Consequently, the level spacing of resonant peaks is $\Delta E = 2\pi\hbar v_F/L \approx \hbar v_F/R$. The result is consistent with Fig. 3b where we obtained $\Delta E \approx \alpha\hbar v_F/R$ and $\alpha = 0.96 \pm 0.06$. For a rectangular GQD, there are two incident angles $\theta_1$ and $\theta_2$ for a quasi-particle incident to the junction boundary (see supplemental Fig. S9e). Because of $\theta_1 + \theta_2 = \frac{\pi}{2}$, these two angles are not independent. In order to obtain a high reflection probability, the incident angle should neither be too large nor too small. Fig. S9d-9f plot several typical closed interference paths under such restriction. Seen as Fig. S9e and S9f, the interference paths cross inside the bulk. Thus, one can expect that the electron wave will constructively or destructively interfere in the interior of the GQD, which leads to the complex patterns in the STS maps (see Fig. 4b and supplemental Fig. S8). In addition, from Fig. S9e and S9f in the supplemental material, we estimate that the length of the closed path will not deviate $2\pi R$ too much. This may be the reason that $\Delta E \approx \alpha\hbar v_F/R$ still holds in the quasi-rectangle GQD and in other GQDs with different geometries, as shown in Fig. 3b.

In summary, we demonstrate that nanoscle and atomically-sharp *p-n* junctions can be created by locally changing the graphene-Cu separations via monolayer-vacancy-islands engineering on Cu substrate during the graphene growth process. Via STM measurements, we directly image the wavefunctions around the *p-n* junctions. Our results can be treated as an atomic-scale verification of the quantum electron optics based on graphene.

**Note added**: During the submission of this work, we became aware of two related works [39,40] showing exotic properties of the quasi-bound states in GQDs in a continuous graphene sheet.

**Acknowledgments**

This work was supported by the National Natural Science Foundation of China (Grant Nos. 11674029, 11422430, 11374035, 11374219, 11504008, 11674028), the Natural Science Foundation of Jiangsu Province, China (Grant No. BK20160007), the National Basic Research Program of China (Grants Nos. 2014CB920903, 2013CBA01603, 2014CB920901), the program for New Century Excellent Talents in University of the Ministry of Education of China (Grant No. NCET-13-0054), China Postdoctoral Science Foundation Funded Project (Grant Nos. 2016M600952). L.H. also acknowledges support from the National Program for Support of Top-notch Young Professionals and support from "the Fundamental Research Funds for the Central Universities".

**Author contributions**

K.K.B. performed the STM experiments. Y.C.W. and K.K.B. synthesized the samples. K.K.B., J.B.Q., and Y.W.L. analyzed the data. J.J.Z. and H.J. performed the theoretical calculations. L.H. conceived and provided advice on the experiment, analysis, and theoretical calculation. L.H. and K.K.B. wrote the paper. All authors participated in the data discussion.

# Reference

[1] Katsnelson, M. I., Novoselov, K. S. & Geim, A. K. Chiral tunnelling and the Klein paradox in graphene. *Nature Phys.* **2**, 620 (2006).

[2] Pereira, J. M., Mlinar, V., Peeters, F. M. & Vasilopoulos, P. Confined states and direction-dependent

transmission in graphene quantum wells. *Phys. Rev. B* **74**, 045424 (2006).

[3] Cheianov, V. V. & Fal'Ko, V. I. Selective transmission of Dirac electrons and ballistic magnetoresistance of n-p junctions in graphene. *Phys. Rev. B* **74**, 041403 (2006).

[4] Young, A. & Kim, P. Quantum transport and Klein tunneling in graphene heterojunctions. *Nature Phys*. **5**, 222 (2008).

[5] Stander, N., Huard, B. & Goldhabergordon, D. Evidence for Klein tunneling in graphene p-n junctions. *Phys. Rev. Lett.* **102**, 026807 (2009).

[6] He, W. Y., Chu, Z. D. & He, L. Chiral tunneling in a twisted graphene bilayer. *Phys. Rev. Lett.* **111**, 066803 (2013).

[7] Cheianov, V. V., Fal'ko, V. & Altshuler, B. L. The Focusing of Electron Flow and a Veselago Lens in Graphene p-n Junctions. *Science* **315**, 1252 (2007).

[8] Miao, F. *et al*. Phase-coherent transport in graphene quantum billiards. *Science* **317**, 1530 (2007).

[9] Zhang, F. M., He, Y. & Chen, X. Guided modes in graphene waveguides. *Appl. Phys. Lett.* **94**, 212105 (2009).

[10] Williams, J. R., Low, T., Lundstrom, M. S. & Marcus, C. M. Gate-controlled guiding of electrons in graphene. *Nature Nanotechnol.* **6**, 222 (2011).

[11] Rickhaus, P. *et al*. Ballistic interferences in suspended graphene. *Nature Commun.* **4**, 2342 (2013).

[12] Taychatanapat, T., Watanabe, K., Taniguchi, T. & Jarillo-Herrero, P. Electrically tunable transverse magnetic focusing in graphene. *Nature Phys.* **9**, 225 (2013).

[13] Lee, G. H., Park, G. H. & Lee, H. J. Observation of negative refraction of Dirac fermions in graphene. *Nature Phys.* **11**, 925 (2015).

[14] Zhao, Y. *et al.* Creating and probing electron whispering-gallery modes in graphene. *Science* **348**,

672 (2015).

[15] Chen, S. *et al.* Electron optics with p-n junctions in ballistic graphene. *Science* **353**, 1522 (2016).

[16] Williams, J. R., DiCarlo, L. & Marcus, C. M. Quantum Hall effect in a gate-controlled p-n junction of graphene. *Science* **317**, 638 (2008).

[17] Lee, J. *et al.* Imaging electrostatically confined Dirac fermions in graphene quantum dots. *Nature Phys.* **12**, 1032 (2016).

[18] Gutierrez, C., Brown, L., Kim, C.-J., Park, J. & Pasupathy, A. N. Klein tunnelling and electron trapping in nanometre-scale graphene quantum dots. *Nature Phys.* **12**, 1069 (2016).

[19] Bai, K. K., Qiao, J. B., Jiang, H., Liu, H. W. & He, L. Massless Dirac Fermions Trapping in a Quasi-one-dimensional npn Junction of a Continuous Graphene Monolayer. *Phys. Rev. B* **95**, 201406(R) (2017).

[20] de la Figuera, J., Prieto, J. E., Ocal, C. & Miranda, R. Creation and motion of vacancy islands on solid surfaces: A direct view. *Solid State Commun.* **89**, 815 (1994).

[21] Hannon, J. B. *et al.* Surface self-difusion by vacancy motion: island ripening on Cu(001). *Phys. Rev. Lett.* **79**, 2506 (1997).

[22] Pai, W. W. *et al.* Evolution of Two-Dimensional Wormlike Nanoclusters on Metal Surfaces. *Phys. Rev. Lett.* **86**, 3088 (2001).

[23] Tapaszto, L. *et al.* Breakdown of continuum mechanics for nanometre-wavelength rippling of graphene. *Nature Phys.* **8**, 739 (2012).

[24] Rasool, H. I. *et al.* Atomic-scale characterization of graphene grown on copper (100) single crystals. *J. Am. Chem. Soc.* **133**, 12536 (2011).

[25] Li, X. *et al.* Large-Area Synthesis of High-Quality and Uniform Graphene Films on Copper Foils.

*Science* **324**, 1312 (2009).

[26] Khomyakov, P. A. *et al*. First-principle study of the interaction and charge transfer between graphene and metals. *Phys. Rev. B* **79**, 175425 (2009).

[27] Zhao, L. *et al*. Visualizing Individual Nitrogen Dopants in Monolayer Graphene. *Science* **333**, 999 (2011).

[28] Zhang, Y. *et al*. Giant phonon-induced conductance in scanning tunnelling spectroscopy of gate-tunable graphene. *Nature Phys.* **4**, 627 (2008).

[29] Zhang, Y., Brar, V. W., Girit, C., Zettl, A. & Crommie, M. F. Origin of spatial charge inhomogeneity in graphene. *Nature Phys.* **5**, 722 (2010).

[30] Yin, L.-J. *et al*. Landau quantization of Dirac fermions in graphene and its multilayers. *Front. Phys.* **12**, 127208 (2017).

[31] Hewageegana, P. & Apalkov, V. Electron localization in graphene quantum dots. *Phys. Rev. B* **77**, 245426 (2008).

[32] Matulis, A. & Peeters, F. M. Quasibound states of quantum dots in single and bilayer graphene. *Phy. Rev. B* **77**, 115423 (2007).

[33] Wu, J. S. & Fogler, M. Scattering of two-dimensional massless Dirac electrons by a circular potential barrier. *Phys. Rev. B* **90**, 235402 (2014).

[34] Bardarson, J. H., Titov, M. & Brouwer, P. W. Electrostatic confinement of electrons in an integrable graphene quantum dot. *Phys. Rev. Lett.* **102**, 226803 (2009).

[35] Downing, C. A., Stone, D. A. & Portnoi, M. E. Zero-energy states in graphene quantum dots and rings. *Phys. Rev.B* **84**, 2149 (2011).

[36] Subramaniam, D. *et al*. Wave-function mapping of graphene quantum dots with soft confinement.

*Phys. Rev. Lett.* **108**, 046801 (2012).

[37] Hamalainen, S. K. *et al*. Quantum-confined electronic states in atomically well-defined graphene nanostructures. *Phys. Rev. Lett.* **107**, 236803 (2011).

[38] Yin, L.-J., Jiang, H., Qiao, J.-B. & He, L. Direct imaging of topological edge states at a bilayer graphene domain wall. *Nature Commun.* **7**, 11760 (2016).

[39] Ghahari, F., et al. An on/off Berry phase switch in circular graphene resonators. *Science* **356**, 845 (2017).

[40] Jiang, Y., et al. Tuning a circular p-n junction in graphene from quantum confinement to optical guiding. arXiv: 1705.07346. To appear in Nature Nano.

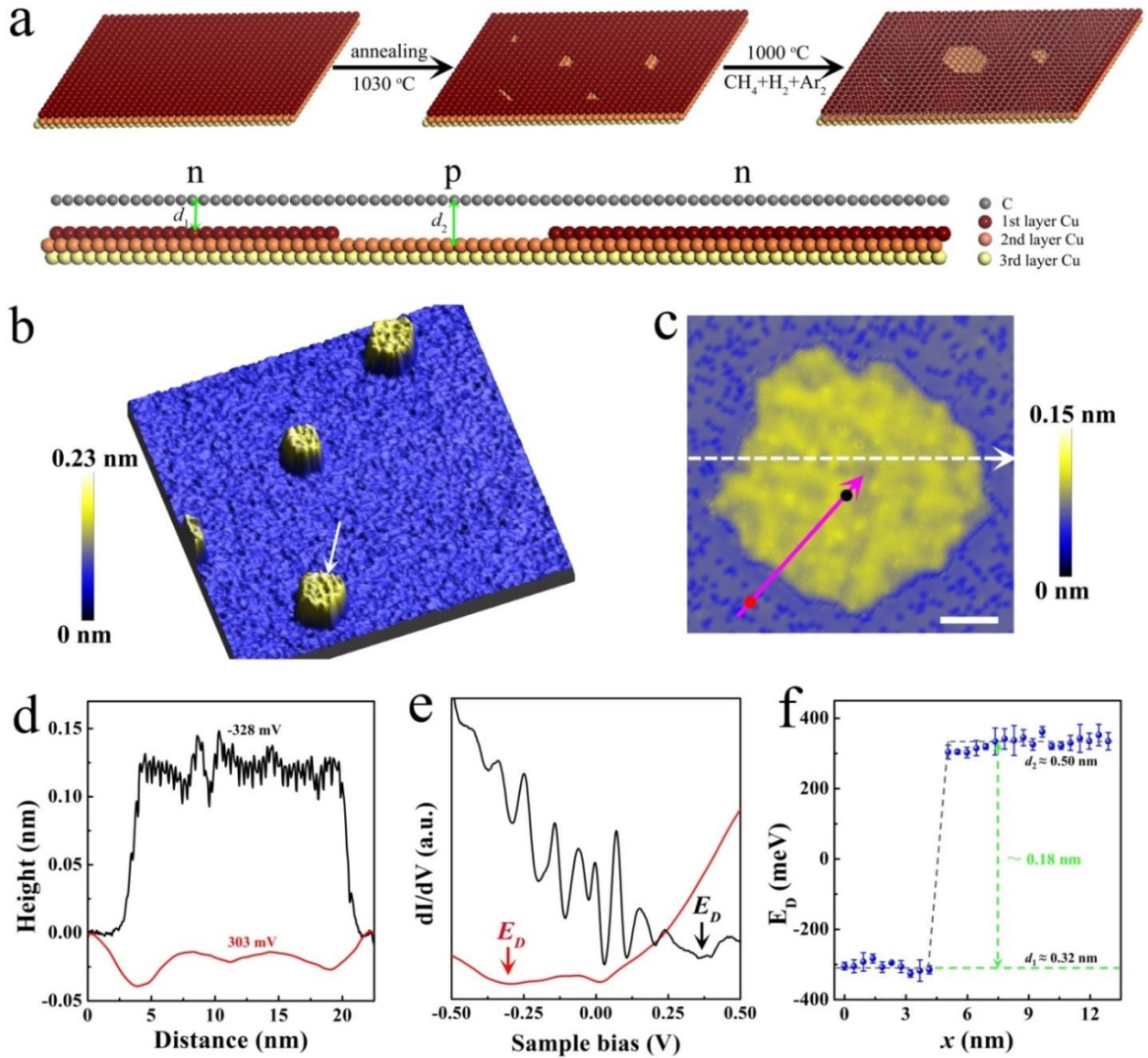

**Figure 1| a, Upper panel**: Schematic model of the formation of monolayer-vacancy-islands on Cu surface during graphene growth process. **Lower panel**: Side view of graphene monolayer on Cu surface around a Cu monolayer-vacancy-island. $d_1$, $d_2$ : graphene-Cu separations in different regions. **n** and **p** denote electron-type and hole-type doping graphene respectively. **b**, (131×122) nm$^2$ three-dimensional STM image of a continuous graphene monolayer on Cu substrate, showing several protrusions. ($V_{sample}$ = -103.2 mV, $I$ = 60 pA). **c**, zoom-in STM image of the protrusion indicated by the white arrow in 1**b**. ($V_{sample}$ = -328 mV, $I$ = 60 pA). Scale bar, 4 nm. **d**, Height profiles recorded at different voltages along the white dashed arrow in 1**c**. **e**, Typical *dI/dV* spectra recorded at the black and red dots in 1**c**. The arrows denote

the position of the Dirac points, $E_D$. For clarity, the curves are offset in $y$-axis. **f,** The figure summarizes $E_D$, obtained at different position along the magenta arrow in **1c**. The gray dashed line is a guide to eye, indicating the average $E_D$ of different region. **d₁** and **d₂** denote the graphene-Cu separations, estimated from the average $E_D$ measured on and off the vacancy-island respectively.

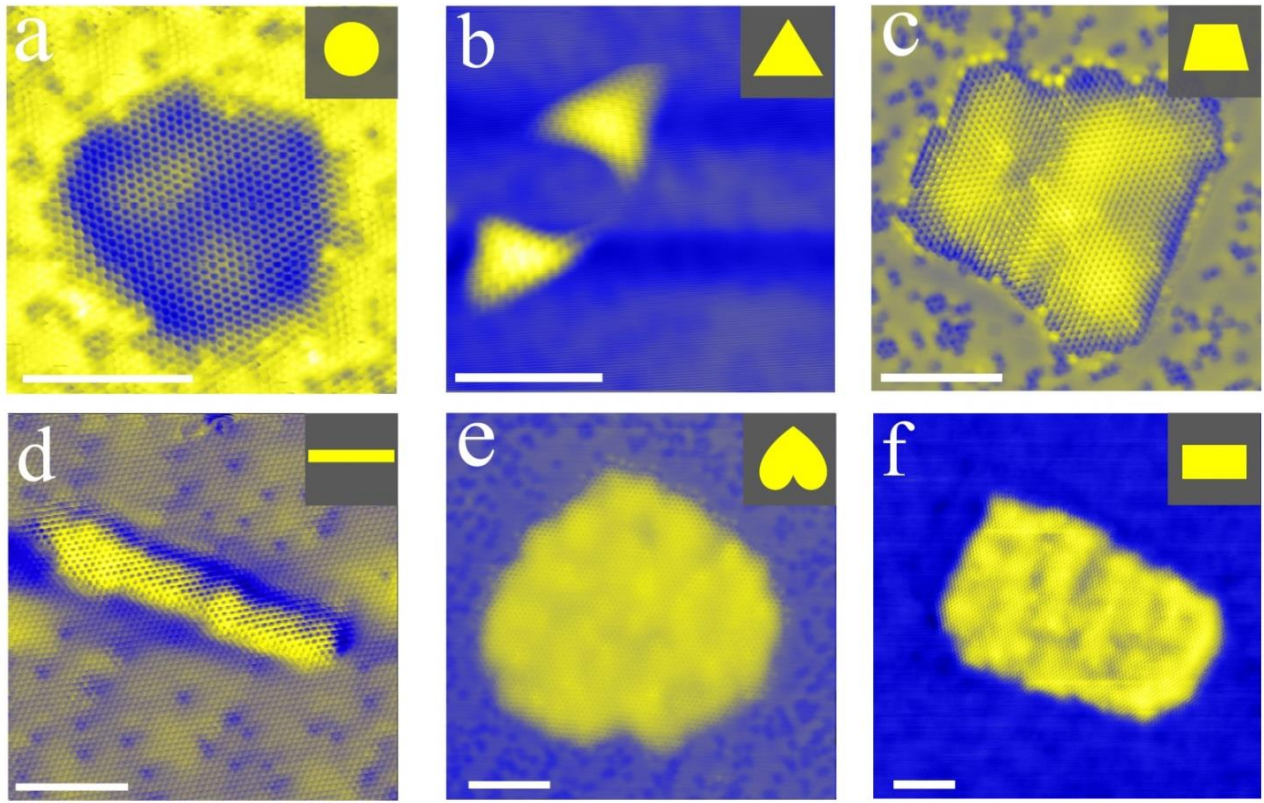

**Figure 2** | Several representative GQDs embedded in a continuous graphene sheet on the Cu substrate Scale bar, 4 nm. Inset: the simplified geometries of GQDs for the main panel. Color scale is from blue to yellow, and blue corresponds to the lower height. **a**: $V_{sample}$ = -481 mV, $I$ = 140 pA; **b**: $V_{sample}$ = -227.5 mV, $I$ = 400 pA; **c**: $V_{sample}$ = -68 mV, $I$ = 100 pA; **d**: $V_{sample}$ = -471 mV, $I$ = 200 pA; **e**: $V_{sample}$ = -494 mV, $I$ = 310 pA; **f**: $V_{sample}$ = -227.5 mV, $I$ = 400 pA.

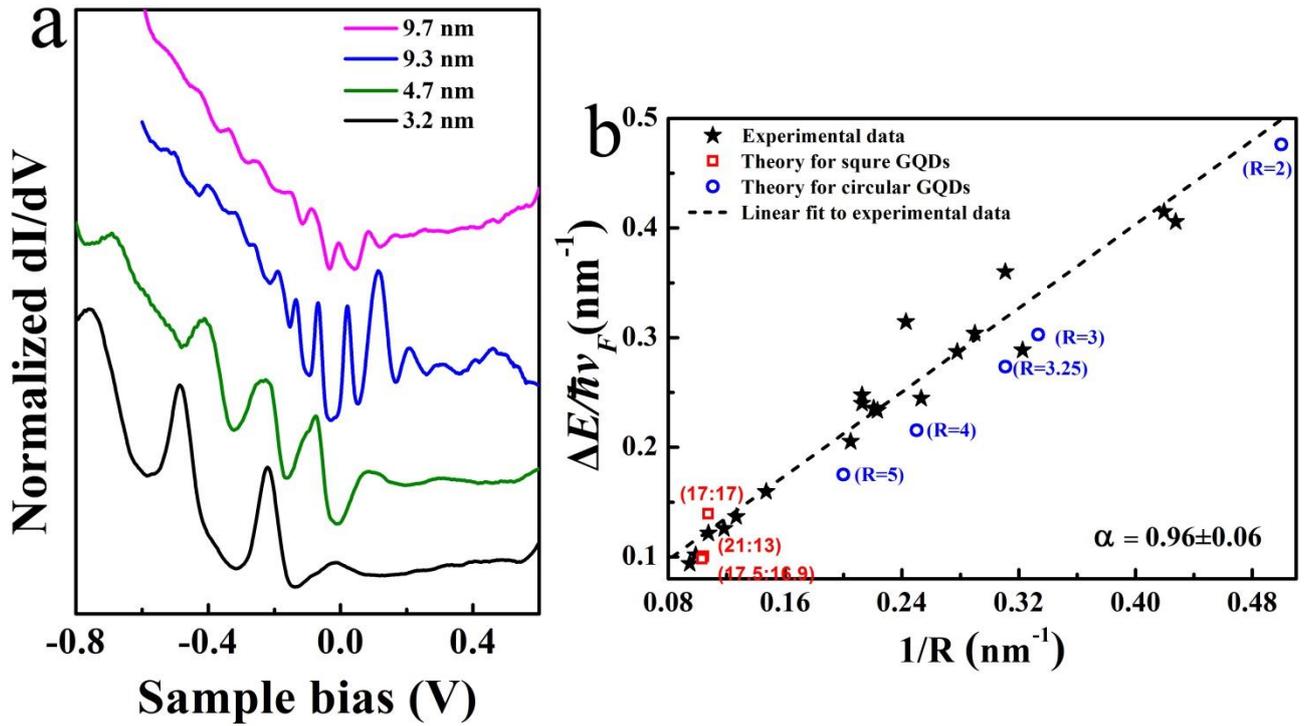

**Figure 3| a,** Normalized *dI/dV* curves for GQDs with different effective radius and geometries. For clarity, the curves are offset in *y*-axis. **b,** Plot of average level spacing for several resonant peaks as a function of inverse effective radius for GQDs. The black stars denote the experimental data, which can be described well by a linear fit: $\Delta E \approx \alpha \hbar v_F / R$ with $\alpha = 0.96 \pm 0.06$. Blue open circle: simulated data for circular GQDs with different radius; Red open square: simulated data for the rectangular GQDs with different aspect ratio.

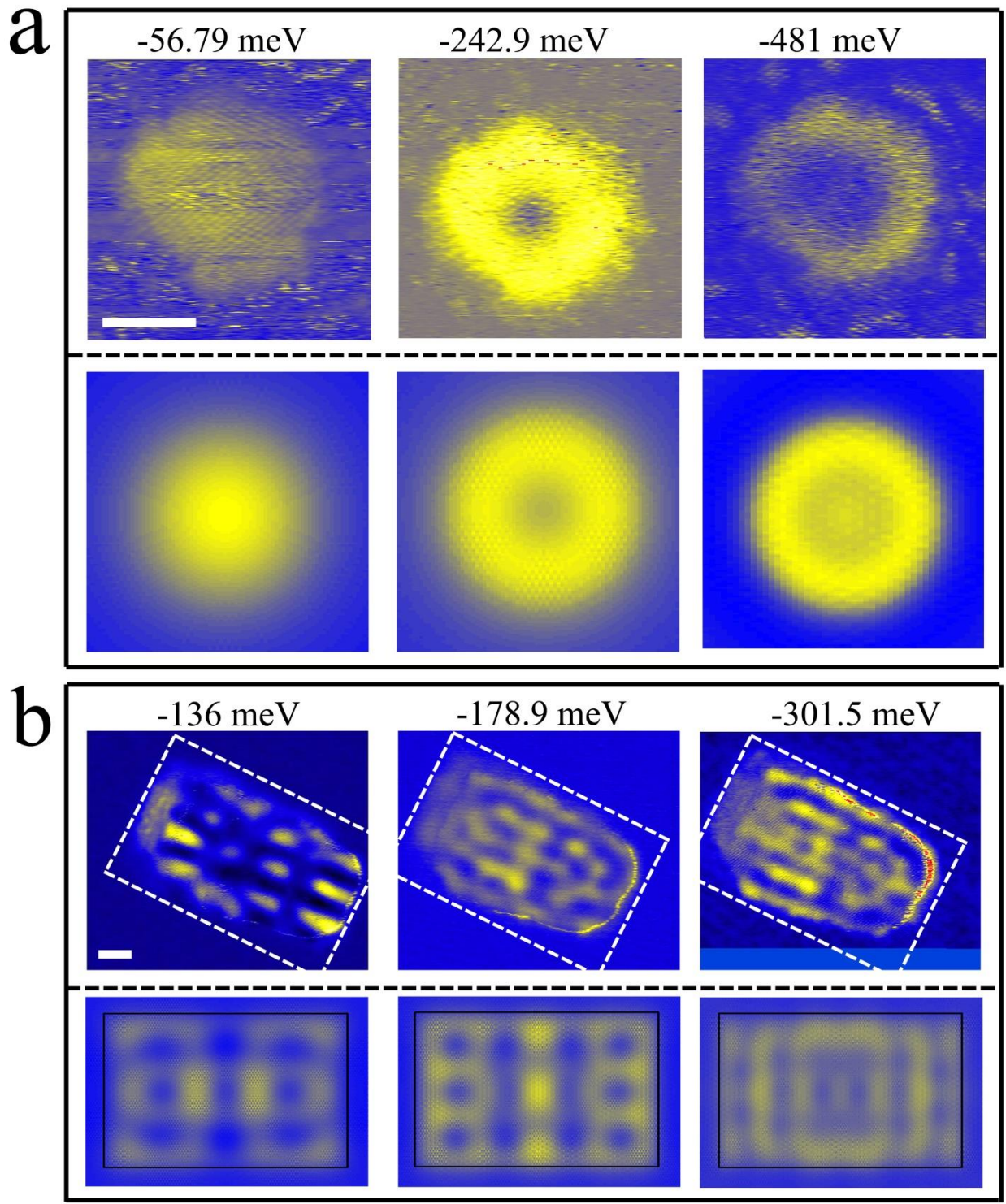

**Figure 4**| Experimental and calculated STS maps recorded at the energy of quasi-bound states for two

representative GQDs. Scale bar, 3 nm. **a**, **Upper panel:** experimental STS maps for a quasi-circular GQD

same as to Fig. **2a**; **Lower panel:** corresponding calculated LDOS maps based on the lattice Green's function method as described in the supplemental material. **b**, **Upper panel**: experimental STS maps for a quasi-rectangular GQD same as to Fig. **2d**. **Lower panel**: corresponding calculated LDOS maps for the region marked by the white dashed frame in the upper panel. Color scale is from blue to yellow, and blue corresponds to the vanishing LDOS.